\documentclass[11pt]{article}
\usepackage{amssymb}
\usepackage{amsmath}
\usepackage[english]{babel}
\usepackage{graphicx}

\def\Journal#1#2#3#4{{#1} {\bf #2}, #3 (#4)}

\def\NCB{\em Nuovo Cim. B}

\def\PLB{{\em Phys. Lett.}  B}
\def\PRL{\em Phys. Rev. Lett.}
\def\PRD{{\em Phys. Rev.} D}

\def\GaC{\em Gravitation and Cosmology}
\def\GaCS{{\em Gravitation and Cosmology} Supplement}

\def\JETPL{\em JETP Lett.}
\def\PAN{\em Phys.Atom.Nucl.}
\def\CQG{\em Class. Quantum Grav.}

\def\MPLA{{\em Mod. Phys. Lett.}  A}
\def\IJTP{\em Int. J. Theor. Phys.}
\def\NJP{\em New J. of Phys.}
\def\JHEP{\em JHEP}

\def\BWP{\em Bled Workshops in Physics}
\def\JPCS{{\em J. Phys.:} Conf. Ser.}

\def\IJMPA{{\em Int. J. Mod. Phys.}  A}

\def\({\left(}
\def\){\right)}

\def\beq{\begin{equation}}
\def\eeq{\end{equation}}
\def\bea{\begin{eqnarray}}
\def\eea{\end{eqnarray}}

\begin{document}

    \begin{center}
        \large \textbf{Probes for 4th generation constituents of dark atoms in Higgs boson studies at the LHC}
    \end{center}

    \begin{center}
   M. Yu. Khlopov$^{1,2,3}$, R. M. Shibaev$^{1}$

    \emph{$^{1}$National Research Nuclear University "Moscow Engineering Physics Institute", 115409 Moscow, Russia \\
    $^{2}$ Centre for Cosmoparticle Physics "Cosmion" 115409 Moscow, Russia \\
$^{3}$ APC laboratory 10, rue Alice Domon et L\'eonie Duquet \\75205
Paris Cedex 13, France}

    \end{center}

\medskip

\begin{abstract}
The nonbaryonic dark matter of the Universe can consist of
new stable charged species, bound in heavy neutral "atoms" by ordinary Coulomb
 interaction. Stable $\bar U$ (anti-$U$)quarks of 4th generation, bound in stable colorless ($\bar U \bar U \bar U $) clusters, are captured by the primordial helium, produced in Big Bang  Nucleosynthesis, thus forming
 neutral "atoms" of O-helium (OHe), a specific nuclear interacting dark matter that can provide solution for the puzzles of direct dark matter searches. However, the existence of the 4th generation quarks and leptons should influence the production and decay rates of Higgs boson and is ruled out by the experimental results of the Higgs boson searches at the LHC, if the Higgs boson coupling to 4th generation fermions with is not suppressed. Here we argue that the difference between the three known quark-lepton families and the 4th family can naturally lead to suppression of this coupling, relating the accelerator test for such a composite dark matter scenario to the detailed study of the production and modes of decay of the 125.5 GeV boson, discovered at the LHC.

\end{abstract}
\section{Introduction}
The cosmological dark matter can
consist of dark atoms, in which new stable charged particles are bound by ordinary Coulomb interaction (see \cite{Levels,Levels1,mpla,DMRev,4QGC} for review and references).
In order to avoid anomalous
isotopes overproduction, stable particles with charge -1 (and
corresponding antiparticles as tera-particles \cite{Glashow}) should be absent \cite{Fargion:2005xz}, so that stable
negatively charged particles should have charge -2 only.

There
exist several types of particle models, in which heavy
stable -2  charged species, $O^{--}$, are predicted:
\begin{itemize}
\item[(a)] AC-leptons, predicted
in the extension of standard model, based on the approach
of almost-commutative geometry \cite{Khlopov:2006dk,5,FKS,bookAC}.
\item[(b)] Technileptons and
anti-technibaryons in the framework of walking technicolor
models (WTC) \cite{KK,Sannino:2004qp,Hong:2004td,Dietrich:2005jn,Dietrich:2005wk,Gudnason:2006ug,Gudnason:2006yj}.
\item[(c)] and, finally, stable "heavy quark clusters" $\bar U \bar U \bar U$ formed by anti-$U$ quark of 4th generation
 \cite{Khlopov:2006dk,Q,I,lom,KPS06,Belotsky:2008se}.
\end{itemize}

All these models also
predict corresponding +2 charge particles. If these positively charged particles remain free in the early Universe,
they can recombine with ordinary electrons in anomalous helium, which is strongly constrained in the
terrestrial matter. Therefore cosmological scenario should provide a  mechanism, which suppresses anomalous helium.
There are two possibilities, requiring two different mechanisms of such suppression:
\begin{itemize}
\item[(i)] The abundance of anomalous helium in the Galaxy may be significant, but in the terrestrial matter
there exists a recombination mechanism suppressing this abundance below experimental upper limits \cite{Khlopov:2006dk,FKS}.
The existence of a new strict U(1) gauge symmetry, causing new Coulomb like long range interaction between charged dark matter particles, is crucial for this mechanism. Therefore the existence of dark radiation in the form of hidden photons is inevitable in this approach.
\item[(ii)] Free positively charged particles are already suppressed in the early Universe and the abundance
of anomalous helium in the Galaxy is negligible \cite{mpla,DMRev,I}.
\end{itemize}
These two possibilities correspond to two different cosmological scenarios of dark atoms. The first one is
realized in the scenario with AC leptons, forming neutral AC atoms \cite{FKS}.
The second assumes charge asymmetric case with the excess of $O^{--}$, which form atom-like states with
primordial helium \cite{mpla,DMRev,I}.

If new stable species belong to non-trivial representations of
electroweak SU(2) group, sphaleron transitions at high temperatures
can provide the relationship between baryon asymmetry and excess of
-2 charge stable species, as it was demonstrated in the case of WTC
\cite{KK,KK2,unesco,iwara}.

 After it is formed
in the Standard Big Bang Nucleosynthesis (SBBN), $^4He$ screens the
$O^{--}$ charged particles in composite $(^4He^{++}O^{--})$ {\it
O-helium} ``atoms'' \cite{I}.

In all the models of O-helium, $O^{--}$ behaves either as lepton or
as specific "heavy quark cluster" with strongly suppressed hadronic
interaction. Therefore O-helium interaction with matter is
determined by nuclear interaction of $He$. These neutral primordial
nuclear interacting objects can explain the modern dark matter
density and represent a nontrivial form of strongly
interacting dark matter \cite{McGuire:2001qj,McGuire1,McGuire2,Starkman,Starkman2,Starkman3,Starkman4,Starkman5,Starkman6}.

The cosmological scenario of O-helium Universe allows to explain many results of experimental searches for dark matter \cite{mpla}. Such scenario is insensitive to the properties of $O^{--}$, since the main features of OHe dark atoms are determined by their nuclear interacting helium shell. It challenges experimental probes for the new stable charged particles at accelerators and such probes strongly depend on the nature of $O^{--}$.

Stable $-2$ charge states ($O^{--}$) can be elementary like AC-leptons or technileptons,
or look like elementary as technibaryons. The latter, composed of techniquarks, reveal their structure at much higher energy scale and should be produced at the LHC as
elementary species. They can also be composite like "heavy quark
clusters" $\bar U \bar U \bar U$ formed by anti-$U$ quark in the models of stable fourth
generation \cite{Q,I}.

In the context of composite dark matter scenario accelerator probe for new stable quark-lepton generation
acquires the meaning of critical test for the existence of
charged constituents of cosmological dark matter.

The signature for double charged AC leptons and techniparticles is unique and distinctive what has already allowed
to obtain the lower bound on their mass of 430 GeV in the ATLAS experiment \cite{2qatlas}.

Test for composite $O^{--}$ at the LHC can be only indirect:
through the search for heavy hadrons, composed of single $U$ or
$\bar U$ and light quarks (similar to R-hadrons) \cite{4QGC}, or by virtual effects of 4th generation fermions in the processes with known particles.
 Here we
study a possibility for experimental probe of the hypothesis of stable 4th generation in the studies of 125.5 GeV Higgs boson, discovered in the ATLAS \cite{atlas,atlas1} and the CMS experiments \cite{cms1,cms} at the LHC. The results of these studies \cite{atlas,atals1,cms1,cms} indicate that the number of the detected events, being the production cross section times the decay rate of Higgs boson to two-photon channel, is consistent with the prediction of the Standard model. On the other hand, as it was first revealed in \cite{nuHiggs} the existence of 4th generation leads to enhancement of the main mechanism of Higgs boson production in $pp$ collisions, what puts constraints on the effect of 4th generation particles and practically excludes the possibility of their full strength coupling to 125.5 GeV Higgs boson.

In the model of stable 4th generation the difference of these fermions from the quarks and leptons of the three known families is related to some new conserved charge (which can be even a gauge charge) that protects the stability the lightest quarks and leptons ($U$-quark and the 4th neutrino $N$). The experimental lower limits on the new quarks and leptons make these particles to be heavier than the three light families, what can be explained by the existence of an additional mechanism of their mass generation, e.g. in the framework of multi-Higgs models. It can naturally lead to suppression of the coupling of 4th generation fermions to the 125.5 GeV Higgs boson, discovered at the LHC. Here we explore a possibility to make the 4th generation hypothesis consistent with the experimental data on the two gamma decays of Higgs boson, what opens the door to the indirect probes of the charged constituents of composite dark matter in the detailed studies of production and modes of decay of the 125.5 GeV Higgs boson.

\section{\label{accelerators} Effects of 4th generation in Higgs boson production and decay}
\subsection{\label{quarks} The stable 4th generation}
%\subsection{\label{4generation} Stable particles of 4th generation matter}
The precision data
on the parameters of the $W$ and $Z$ bosons of the Standard model do not exclude \cite{Maltoni:1999ta,Maltoni2,Maltoni3,Maltoni4} the existence of
the  4th generation of quarks and leptons.

The existence of the 4th generation can follow from the heterotic string phenomenology and
its difference from the three known light generations can be
explained by a new conserved charge, possessed only by
its quarks and leptons \cite{Q,I,lom,KPS06,Belotsky:2008se,Belotsky:2000ra,Belotsky:2005uj,Belotsky:2004st}. Strict conservation of this charge makes the
lightest particle of 4th family (the 4th neutrino) absolutely
stable, but it was shown in \cite{Belotsky:2000ra,Belotsky:2005uj,Belotsky:2004st} that this neutrino cannot be the dominant form of the dark matter.
The same conservation law requires the lightest quark to be long living
\cite{Q,I}. In principle the lifetime of $U$ can exceed the age of the
Universe, if $m_U<m_D$ \cite{Q,I}.

 Due to their Coulomb-like QCD attraction ($\propto \alpha_{c}^2 \cdot m_U$, where $\alpha_{c}$ is the QCD constant) stable double and triple $U$ bound states $(UUq)$, $(UUU)$  can exist
\cite{Q,Glashow,Fargion:2005xz,I,lom,KPS06,Belotsky:2008se}. The corresponding antiparticles can be formed by heavy antiquark $\bar U$. Formation of these double and
triple states at accelerators and in
cosmic rays is strongly suppressed by phase space constraints, but they can be formed in early
Universe and strongly influence cosmological evolution of 4th
generation hadrons. As shown in \cite{I}, \underline{an}ti-
\underline{U}-\underline{t}riple state called \underline{anut}ium
or $\Delta^{--}_{3 \bar U}$ is of a special interest. This stable
anti-$\Delta$-isobar, composed of $\bar U$ antiquarks can be bound with $^4He$ in atom-like state
of O-helium \cite{Khlopov:2006dk}.

Since simultaneous production of three $U \bar U$ pairs and
their conversion in two doubly charged quark clusters $UUU$
is suppressed, the only possibility to test the
models of composite dark matter from 4th generation in the collider experiments is a search for production of stable hadrons containing single $U$ or $\bar U$ \cite{4QGC} or for effects of 4th generation quarks and leptons in the processes with known particles. Such effects should influence the production and decay rates of the Higgs boson, so the first step in such analysis is to check the consistency of the 4th generation hypothesis with the experimental data on the 125.5 GeV Higgs boson, discovered at the LHC. In the present paper we show that the suppression of couplings of this boson to the 4th generation quarks and leptons can make compatible the experimental data on the two photon decay mode of the Higgs boson with this hypothesis. The precise determination of the range of such couplings implies the account for the two-loop electroweak corrections and for the results of the Higgs searches in other decay channels what goes beyond the scope of the present paper.

We take for definiteness the masses of the fourth generation $U$ and $D$ quarks to be about 350GeV, of the lepton $E$ about 100 GeV, the mass of the 4th neutrino $N$ about 50 GeV (see \cite{DMRev} for the recent review and references), and vary the coupling constants of the fourth generation fermions to the 125.5 GeV Higgs boson taking into account their possible suppression.

\subsection{\label{production} Effects of 4th generation in Higgs boson production}
\begin{figure}
\begin{center}
\includegraphics[scale=1.3]{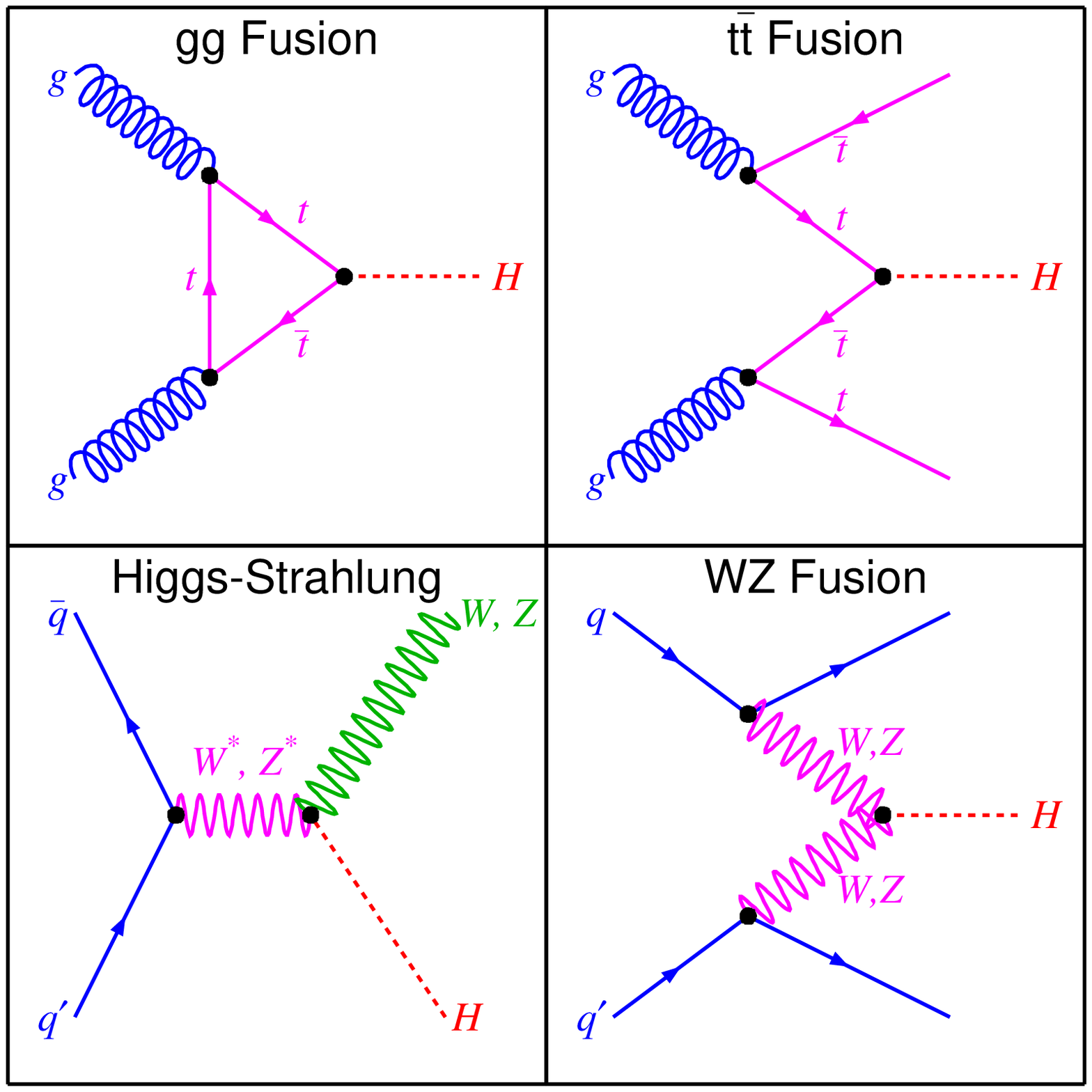}
\end{center}
\caption{Feynman diagrams of the main processes of Higgs boson production in $pp$ collisions (taken from \cite{wiki}).}
\label{Diag}
\end{figure}

\begin{figure}
\begin{center}
\includegraphics[scale=0.7]{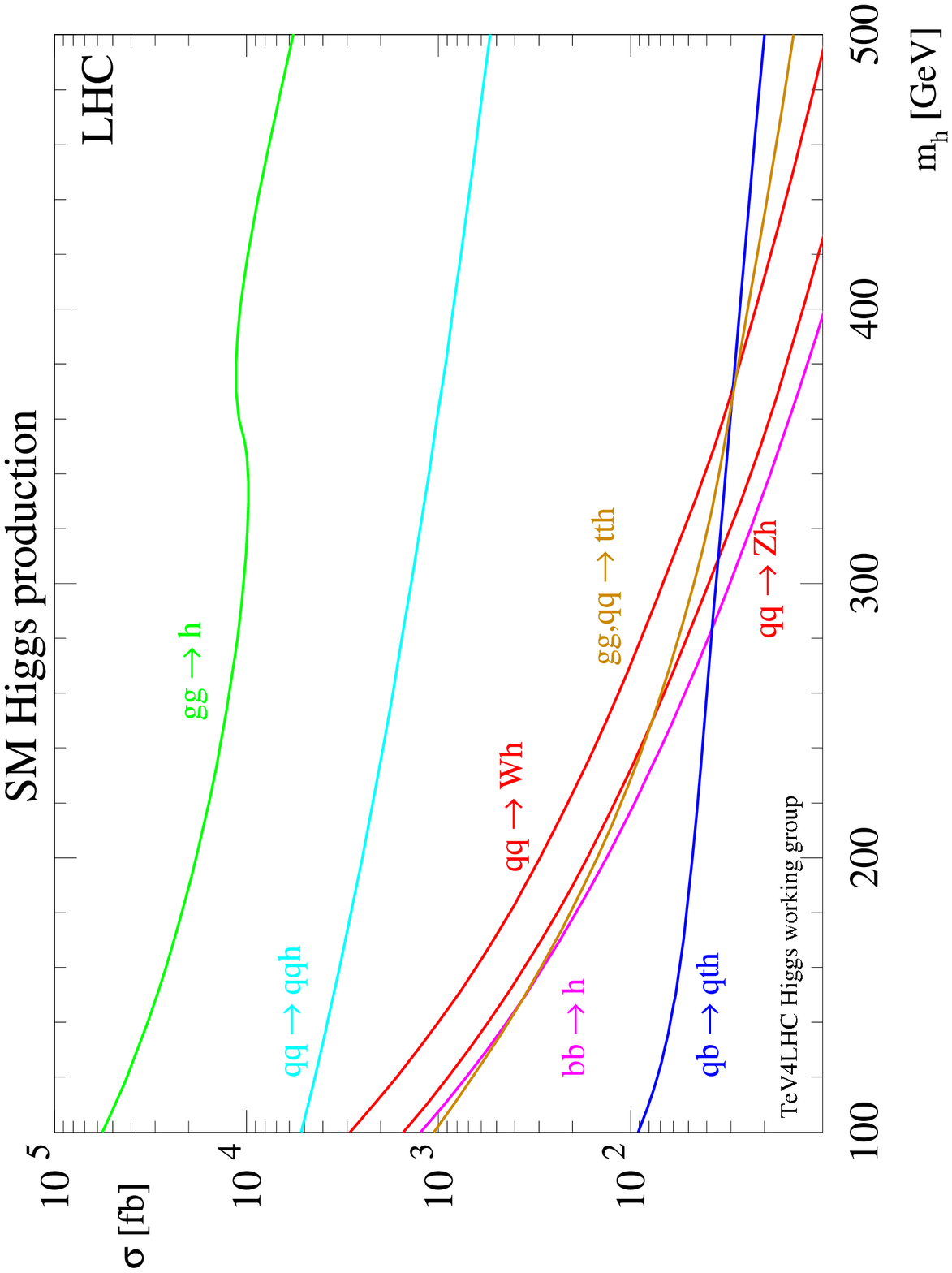}
\end{center}
\caption{The cross section of the main processes of Higgs boson production in $pp$ collisions at LHC (from \cite{SMH}).}
\label{SMHiggsP}
\end{figure}

\begin{figure}
\begin{center}
\includegraphics[scale=0.7]{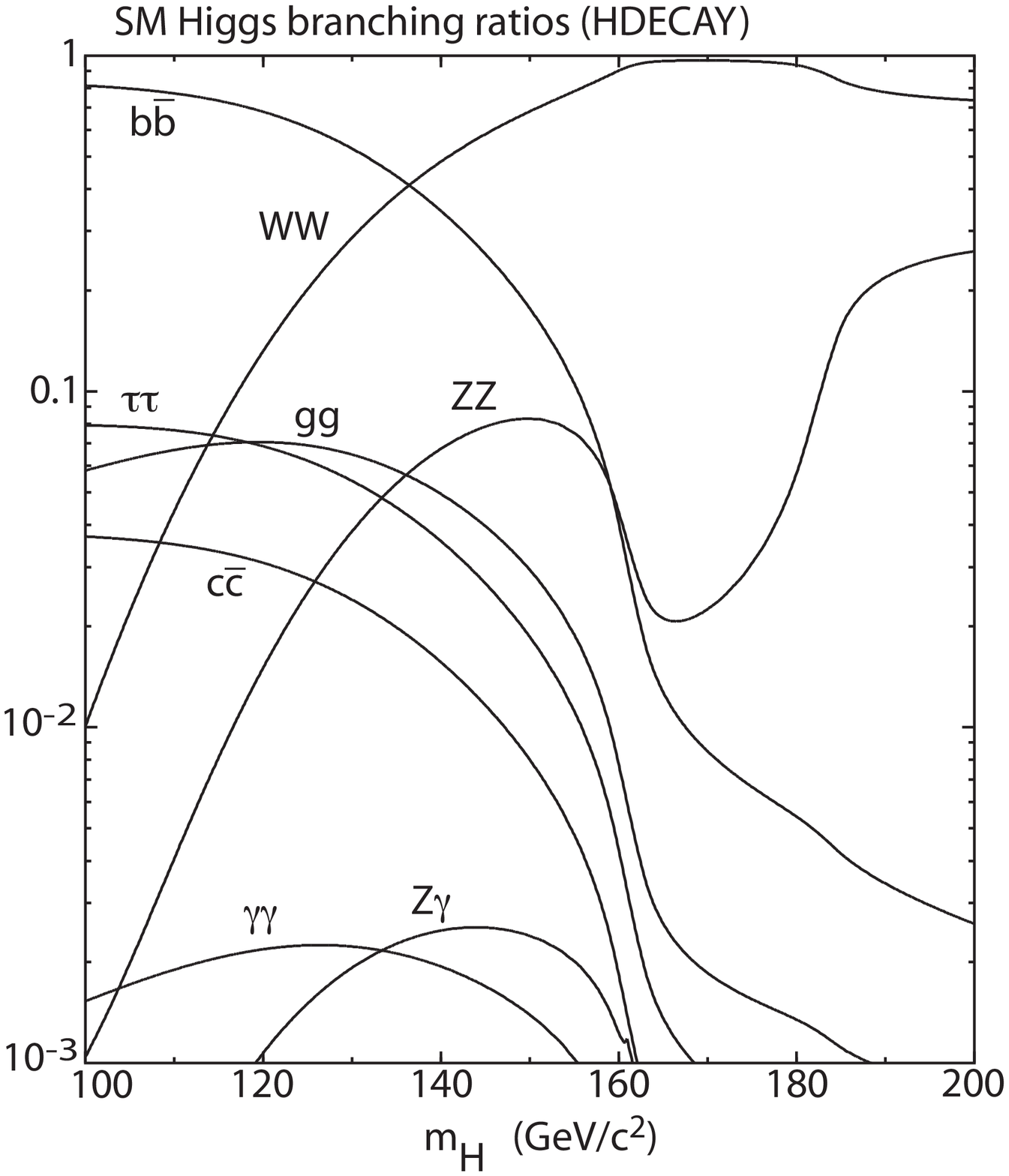}
\end{center}
\caption{The branching ratio of Higgs boson decay modes in the Standard model(from \cite{SMH}).}
\label{SMHiggsD}
\end{figure}

The main channels of Higgs boson production in $pp$ collisions are presented on Fig. \ref{Diag}.
Their relative contribution was calculated in \cite{SMH} and presented on Figs. \ref{SMHiggsP} and \ref{SMHiggsD}.

The dominant mechanism of Higgs production, related with the gluon-gluon fusion, goes through the loop with virtual quarks. Thus the existence of the quarks of fourth generation should influence the amplitude of this process. The corresponding contribution depends on the mass of these quarks and on their Yukawa coupling to the Higgs boson. The value of this coupling constant can vary due to the fact that the large mass of the fourth generation fermions can result from a combined effect of several Higgs fields so that the coupling to the 125.5 GeV Higgs boson can be suppressed. In the present work we calculate the cross section of the Higgs boson production with the use of the HIGLU program \cite{higlu} for different parameters of the model of 4th generation with the account for the QCD corrections.

%(\ref{crossec}).
\subsection{\label{decay} Effects of 4th generation in two-photon Higgs boson decay}
\begin{figure}
\begin{center}
\includegraphics[scale=0.6]{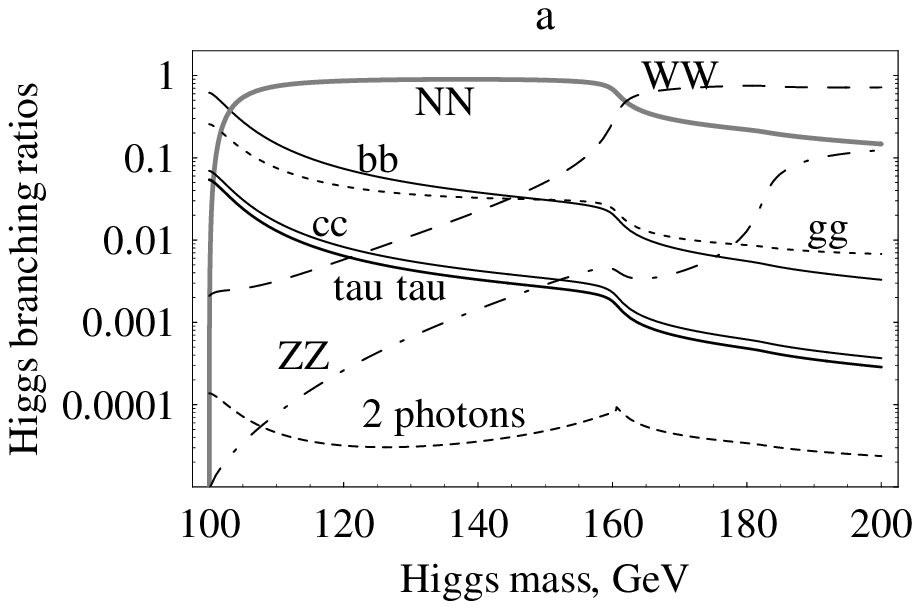}
\end{center}
\caption{Branching ratios of the main processes of Higgs boson decay in the model of stable 4th generation (from \cite{nuHiggs}).}
\label{4fHd}
\end{figure}

%\begin{figure}
%\begin{center}
%\includegraphics[scale=0.5]{Xsection.eps}
%\end{center}
%\caption{Cross sections of production of 4th generation particles (N, E, U (D)) at LHC. Solid and dashed curves correspond to c.m. energies 7 %and 14 TeV respectively.
%Horizontal dashed line shows approximate level of sensitivity to be reached in 2012 (at the energy 7).}
%\label{crossec}
%\end{figure}
Effects of 4th generation in the decay rates of Higgs boson, calculated in \cite{nuHiggs}, are presented on Fig. \ref{4fHd}.
The two-photon decay, considered here, goes through a loop, in which contribute all the charged particles. The width of this decay, calculated as in the Standard model \cite{Okunbook} and modified with the account for the suppression of Higgs coupling constants, is given by
\beq
{\Gamma}(H \to \gamma \gamma) = {\frac{G_F {\alpha}^2 {m_H}^3}{128\sqrt{2} {\pi}^3}}{\left| \sum_{i=W,f} {g_i} {N_c} {e_i}^2 {F_i}\right|}^2,
\label{width}
\eeq
where $\alpha_s \approx 0.118$ and $\alpha \approx 1/137$ are the QCD and QED running gauge constants, $N_c$ and $e_i$ are the number of color degrees of freedom and the electric charge of an $i$-th particle, respectively. The loop factor $F_i$ is given by
 $$
F_{boson} = 2+3\delta+3\delta(2-\delta)f(\delta)
    $$
for bosons and
    $$
F_{fermion} = -2{\delta}(1+(1-\delta)f(\delta))
    $$
for fermions, where
    $$
f(\delta)=\left\{
\begin{aligned}
arctg{\frac{1}{\sqrt{\delta-1}}}, \delta &\ge& 1 \\
\frac{i}{2} \ln \frac{1+\sqrt{1-\delta}}{1-\sqrt{1-\delta}} + \frac{\pi}{2},\delta &<& 1
\end{aligned}
\right.
    $$
The parameter $\delta$ is defined as
  $$
\delta = (2m_i/m_H)^2.
 $$
In the limit of $m_i \gg m_H $ the loop factor depends very weakly on the mass of an intermediate particle, so that the width Eq. (\ref{width}) depends weakly on the mass of the heavy 4th generation fermions. In the Eq. (\ref{width}) $g_i$ are the suppression factors of Higgs boson couplings. They are equal to 1 for the fermions of the three known generations and can vary from 0 to 1 for the quarks and charged lepton of the 4th generation.
\section{The results of calculations}
In the present work we have calculated the ratio
\beq
R = \frac{\sigma_{SM4}(gg \to H)\Gamma_{SM4}(H \to \gamma \gamma)}{\sigma_{SM}(gg \to H)\Gamma_{SM}(H \to \gamma \gamma}
\label{R}
\eeq
of the number of events of two-photon decays of Higgs boson in the model of 4th generation and in the Standard model and compared it with the results of ATLAS \cite{atlas,atlas1} and CMS experiments \cite{cms1,cms} at the LHC. These results within the experimental errors are consistent with the prediction of the Standard model ($R=1$) with the median values $R=1.1$ in CMS and $R=1.3$ in ATLAS that may favor some modest excess of the number of events relative to this prediction. One cannot make compatible the prediction of the model of the 4th generation with these results without suppression of Higgs boson coupling to new quarks and leptons.

The following possibilities for the suppression of the Higgs boson couplings were considered:
\begin{itemize}
\item[(a)] The suppression factor is the same for all the quarks and leptons of the 4th generation.
\item[(b)] The up-type and down-type fermions of the 4th generation have different suppression factors.
\item[(c)] Quarks and leptons of the 4th generation have different suppression factors.
\item[(d)] All the fermions of the 4th generation have different suppression factors.
\end{itemize}

\begin{figure}
\begin{center}
\includegraphics[scale=0.5]{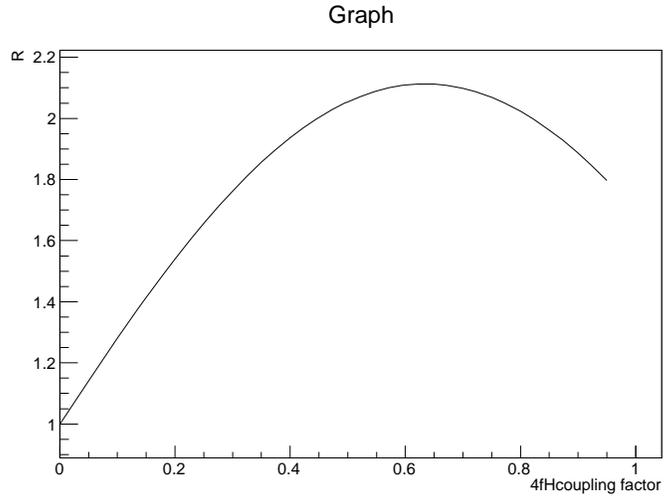}
\end{center}
\caption{The dependence of the ratio $R$ on the suppression factor, which is the same for all the quarks and leptons of the 4th generation.}
\label{Common}
\end{figure}

\begin{figure}
\begin{center}
\includegraphics[scale=0.5]{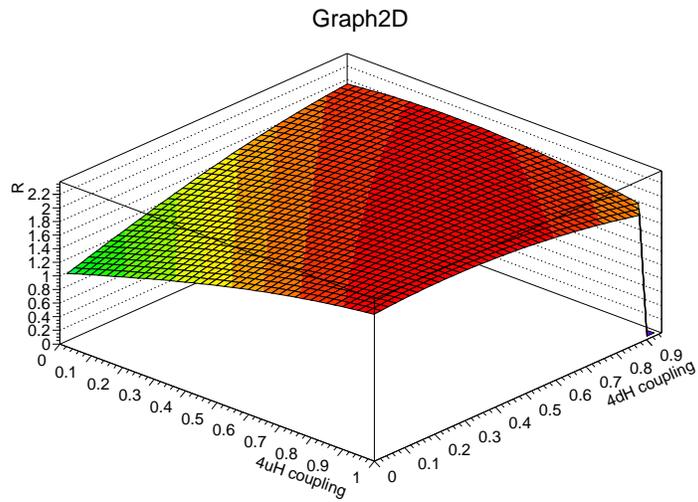}
\end{center}
\caption{The dependence of the ratio $R$ on the suppression factors $4uH$ and $4dH$ of Higgs coupling to correspondingly up- and down- type fermions of the 4th generation.}
\label{ud}
\end{figure}

\begin{figure}
\begin{center}
\includegraphics[scale=0.5]{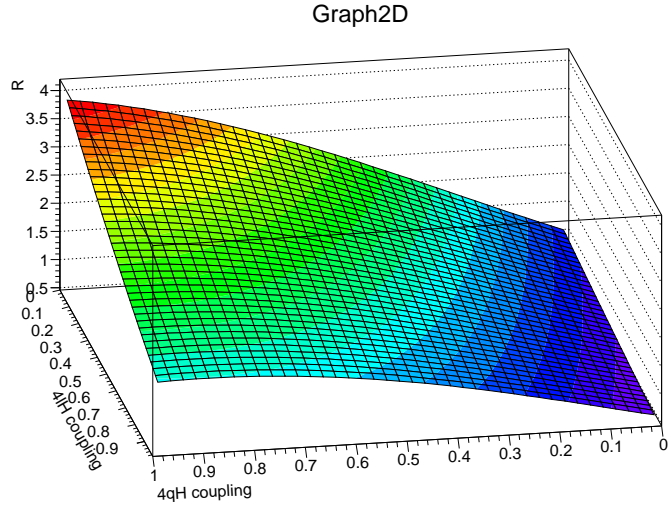}
\end{center}
\caption{The dependence of the ratio $R$ on the suppression factors $4qH$ and $4lH$ of Higgs coupling to correspondingly quarks and leptons of the 4th generation.}
\label{ql}
\end{figure}

\begin{figure}
\begin{center}
\includegraphics[scale=0.5]{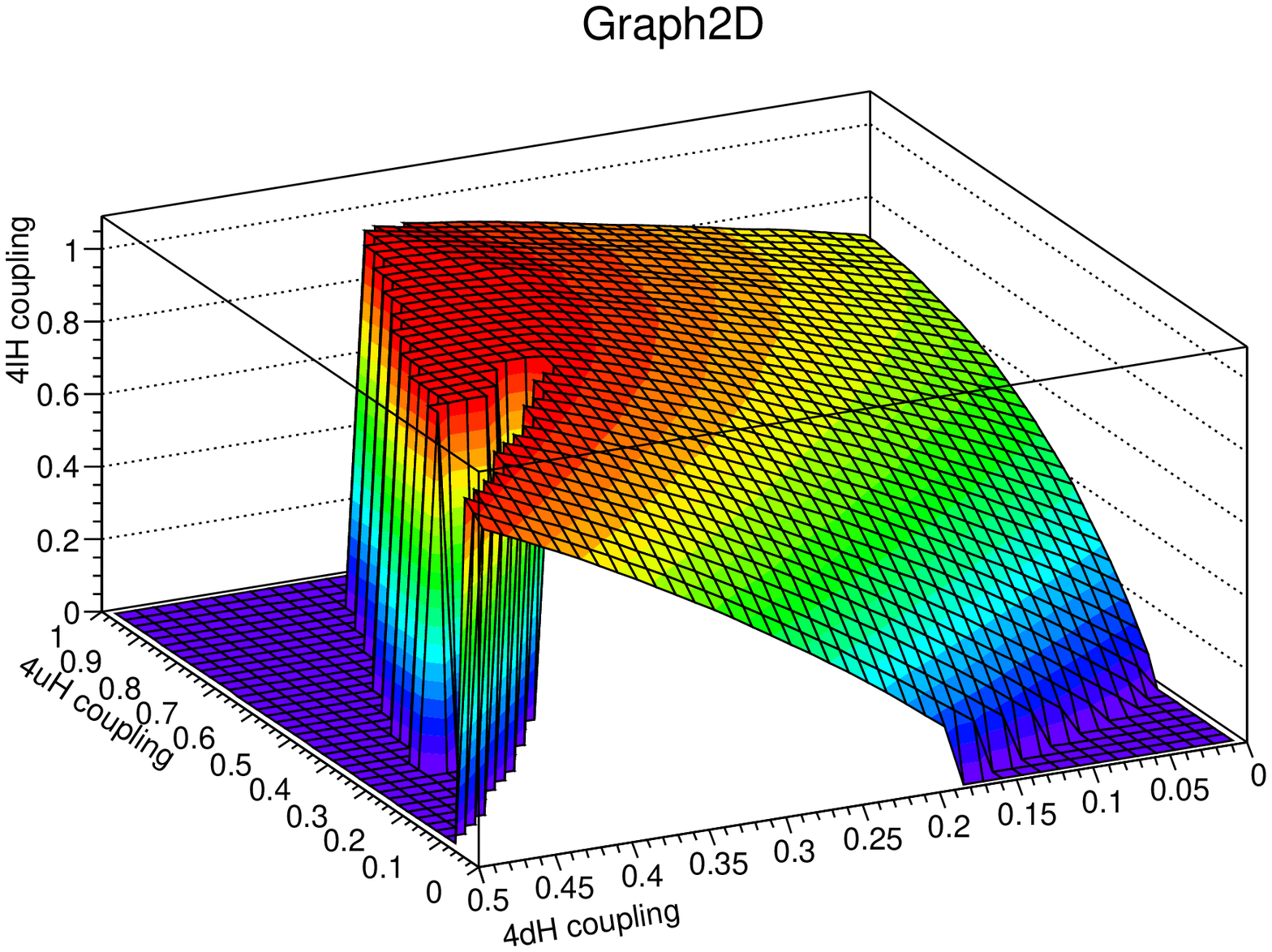}
\end{center}
\caption{The surface in the space of parameters of the suppression factors $4dH$, $4uH$ and $4lH$ in the Higgs coupling to the quarks and leptons of the 4th generation, at which the median value of ratio $R=1.3$, measured in the ATLAS experiment at LHC \cite{atlas}, is reproduced.}
\label{eps13}
\end{figure}

\begin{figure}
\begin{center}
\includegraphics[scale=0.5]{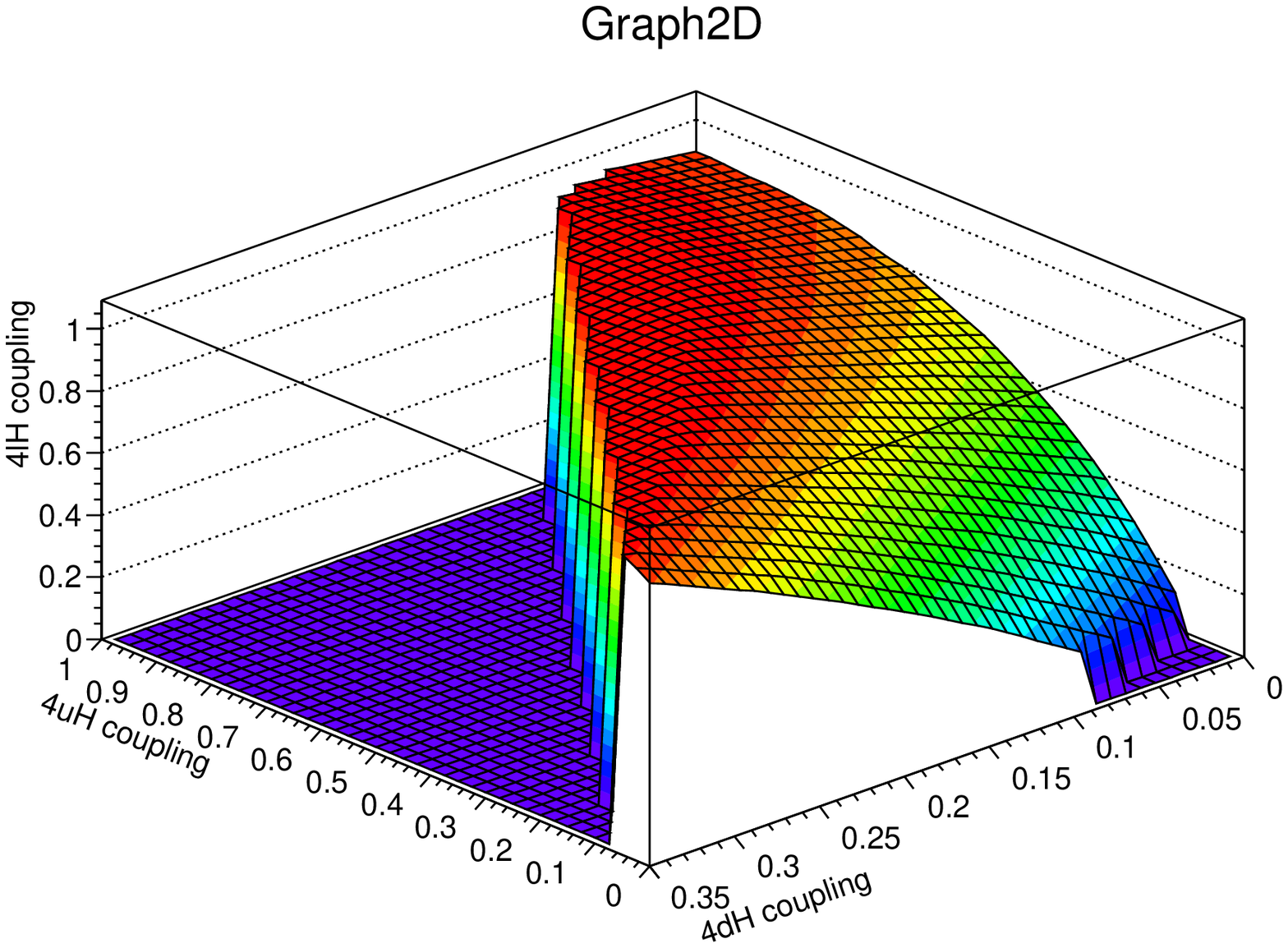}
\end{center}
\caption{The surface in the space of parameters of the suppression factors $4dH$, $4uH$ and $4lH$ in the Higgs coupling to the quarks and leptons of the 4th generation, at which the median value of ratio $R=1.1$, measured in the CMS experiment at LHC \cite{cms}, is reproduced.}
\label{eps11}
\end{figure}

\begin{figure}
\begin{center}
\includegraphics[scale=0.5]{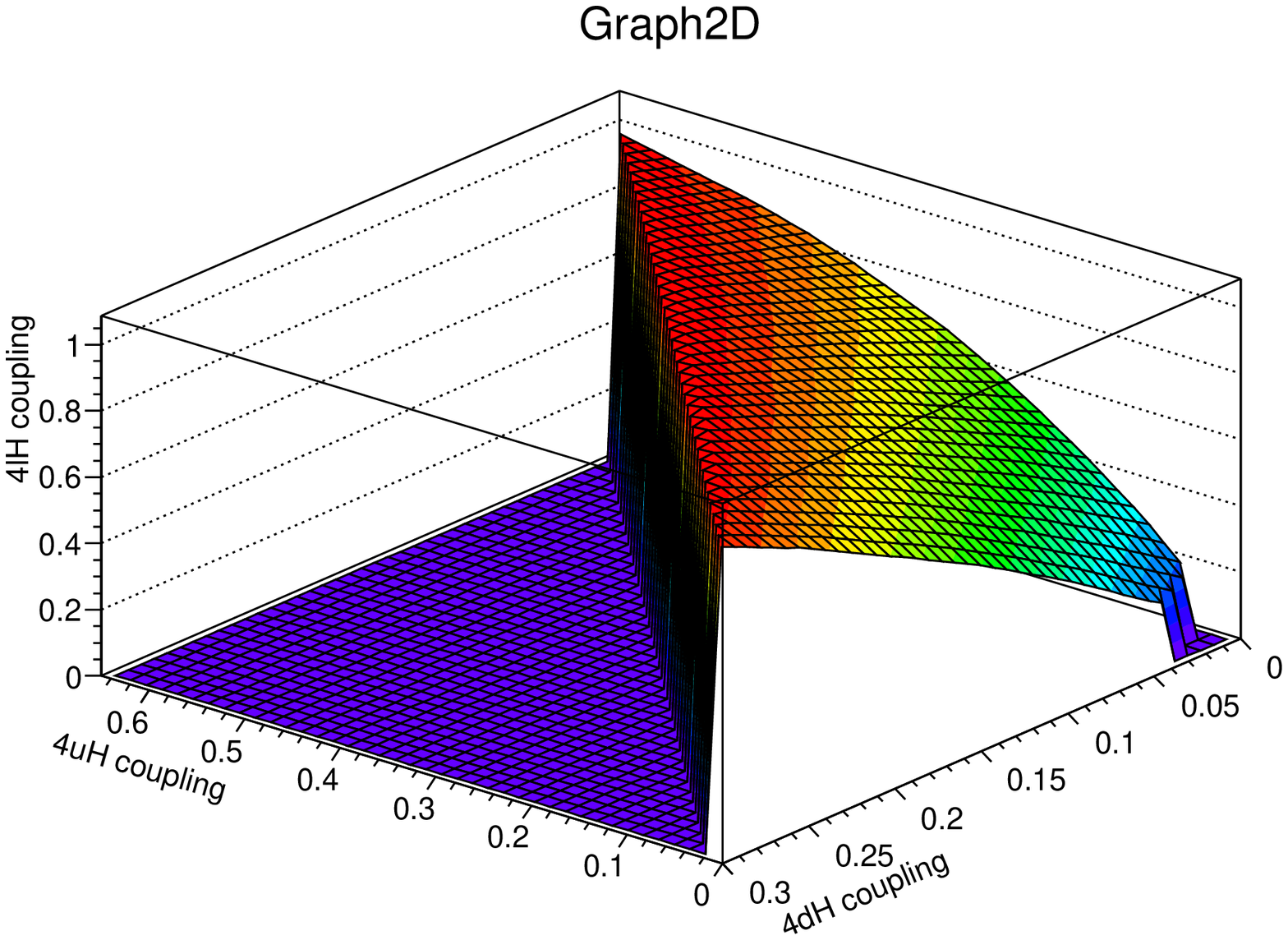}
\end{center}
\caption{The surface in the space of parameters of the suppression factors $4dH$, $4uH$ and $4lH$ in the Higgs coupling to the quarks and leptons of the 4th generation, at which the value of ratio $R=1$, corresponding to the Standard model prediction, is reproduced.}
\label{eps1}
\end{figure}

The results of calculations are presented on Figs. \ref{Common} - \ref{eps1}. The predicted number of events is strongly enhanced by the effects of 4th generation quarks in gluon fusion mechanism of Higgs boson production, so that the corresponding suppression of the Higgs couplings is necessary to make the number of events close to the prediction of the Standard model. However, in the case of different suppression factors for different types of 4th generation particles a nontrivial range of these parameters is possible. One can expect that the analysis of the predictions of the model of the 4th generation for the number of events of other decay modes of the 125.5 GeV Higgs boson will provide an over-determined system of equations for these parameters, making possible the complete test of this model.

\section{Conclusions}

%\medskip
The cosmological dark matter can be formed by
stable heavy double charged particles bound in neutral OHe dark atoms with primordial He nuclei by ordinary Coulomb interaction. This scenario sheds new light on the nature of dark matter and offers nontrivial solution for the puzzles of direct dark matter searches. It can be realized in the model of stable 4th generation and challenges for its experimental probe at accelerators.
 In the context of this scenario search for effects of new heavy quarks and leptons acquires the meaning of direct experimental probe for charged constituents of dark atoms of dark matter.

The $O^{--}$ constituents of OHe in the model of stable 4th generation are "heavy antiquark clusters" $\bar U \bar U \bar U$. Production of such clusters (and their antiparticles) at accelerators is virtually impossible. Therefore experimental test of the hypothesis of stable 4th generation is reduced to the search for stable or metastable $U$ hadrons, containing only single heavy quark or antiquark, or to the studies of virtual effects of 4th generation quarks and leptons in the processes with known particles.

The discovery of the 125.5 GeV Higgs boson at the LHC opens the new room for such indirect test of the model of stable 4th generation. The number of detected events of decays of this boson to the two-photon channel is consistent within the experimental errors with the prediction of the Standard model, putting severe constraints on the contribution of new quarks and leptons. On the other hand, the existence of heavy quarks of the 4th generation should lead to enhancement of the gluon fusion mechanism of Higgs boson production, which is its dominant production mechanism in $pp$ collisions. Therefore, to be compatible with the experimental data the model of the stable 4th generation should involve a mechanism of suppression of new quark and lepton couplings to the 125.5 GeV Higgs boson. Taking into account the difference of the 4th generation from the three known families of quarks and leptons and in particular the lower limits on the masses of new quarks and leptons, one can assume some additional mechanisms of their mass generation, what can lead to suppression of their couplings to the 125.5 GeV Higgs boson.

In the present work we have shown that, indeed, the suppression in the Higgs boson couplings to 4th generation quarks and lepton the can make compatible the existence of this generation with the experimental data on the two-photon decays of the 125.5 GeV Higgs boson. We consider this result as the first step in the thorough investigation of the predictions of the model of stable 4th generation for the whole set of decay channels of Higgs boson. The confrontation of these predictions with the detailed experimental study of the 125.5 GeV Higgs boson will provide the complete test for the composite dark matter scenarios based on the model of the stable 4th generation.

%\bigskip

%{\centering{ \large \textbf{Acknowledgments}} }

\section {Acknowledgments}

%\medskip

We would like to thank K.M. Belotsky for discussions.

%\bigskip

%{\centering{ \large \textbf{References}} }

%\section*{References}

%\medskip

\end{document}